\documentclass{llncs}
\usepackage{llncsdoc}

\usepackage{times}
\usepackage{graphicx}
\usepackage{color}
\usepackage{xspace}
\usepackage{array}
\usepackage{amsmath,  amssymb}
\usepackage{enumerate}
\usepackage{subfig}
\usepackage{multirow}
\usepackage{tabularx}
\usepackage{url}

\begin{document}

\title{Modeling Life as Cognitive Info-Computation}

\author{Gordana Dodig-Crnkovic}
\institute{School of Innovation, Design and Engineering, M\"{a}lardalen University, V\"{a}ster{\aa}s, Sweden\\
\email\small{{gordana.dodig-crnkovic@mdh.se}}
}

 \maketitle
\thispagestyle{empty}

\begin{abstract}
This article presents a naturalist approach to cognition understood as a network of info-computational, autopoietic processes in living systems. It provides a conceptual framework for the unified view of cognition as evolved from the simplest to the most complex organisms, based on new empirical and theoretical results. It addresses three fundamental questions: \textit{what cognition is, how cognition works and what cognition does at different levels of complexity of living organisms}. By explicating the info-computational character of cognition, its evolution, agent-dependency and generative mechanisms we can better understand its life-sustaining and life-propagating role. The info-computational approach contributes to rethinking cognition as a process of natural computation in living beings that can be applied for cognitive computation in artificial systems.
\end{abstract}

\section{Introduction}
\label{sec:Introduction}

It is a remarkable fact that even after half a century of research in cognitive science, cognition still lacks a commonly accepted definition \cite{Lyon_2005}. E.g. NeisserÕs description of cognition as ``all the processes by which sensory input is transformed, reduced, elaborated, stored, recovered and used''~\cite{Neisser_1967} is so broad that it includes present day robots. On the other hand, the Oxford dictionary definition: ``the mental action or process of acquiring knowledge and understanding through thought, experience, and the senses'' applies only to humans. Currently the field of cognitive robotics is being developed where we can learn by construction \textit{what cognition might be} and then, returning to cognitive systems in nature find out what solutions nature has evolved. The process of two-way learning ~\cite{Rozenberg_Kari_2008} starts from nature by reverse engineering existing cognitive agents, while simultaneously trying to design cognitive computational artifacts. We have a lot to learn from natural systems about how to engineer cognitive computers.~\cite{Modha_Ananthanarayanan_Esser_Ndirango_Sherbondy_Singh_2011}

Until recently only humans were commonly accepted as cognitive agents (\textit{anthropogenic} approach in Lyon). Some were ready to ascribe certain cognitive capacities to all apes, and some perhaps to all mammals. The lowest level cognition for those with the broadest view of cognition included all organisms with nervous system. Only a few were prepared to go below that level. Among those very few, the first who were ready to acknowledge a cognitive agency of organisms without nervous system were Maturana and Varela \cite{Maturana_Varela_1980}\cite{Maturana_1970}, who argued that \textit{cognition and life are identical processes}. LyonÕs classification, besides describing the anthropogenic approach, includes a \textit{biogenic} approach based on self-organizing complex systems and autopoiesis. The adoption in the present paper of the biogenic approach through the definition of Maturana and Varela is motivated by the wish to provide a theory that includes all living organisms and artificial cognitive agents within the same framework. 

\section{The Computing Nature, Computational Naturalism and Minimal Cognition}
\label{sec:The Computing Nature, Computational Naturalism and Minimal Cognition}

Naturalism is the view that \textit{nature is the only reality}. It describes nature through its structures, processes and relationships using a scientific approach. Naturalism studies the evolution of the entire natural world, including the life and development of humanity as a part of nature. Computational naturalism (pancomputationalism, naturalist computationalism) is the view that the nature is a huge network of computational processes which, according to physical laws, computes (dynamically develops) its own next state from the current one. Representatives of this approach are Zuse, Fredkin, Wolfram, Chaitin and Lloyd, who proposed different varieties of computational naturalism. According to the idea of computing nature, one can view the time development (dynamics) of physical states in nature as information processing (natural computation). Such processes include self-assembly, self-organization, developmental processes, gene regulation networks, gene assembly, protein-protein interaction networks, biological transport networks, social computing, evolution and similar processes of morphogenesis (creation of form). The idea of computing nature and the relationships between two basic concepts of information and computation are explored in~\cite{Dodig-Crnkovic_Giovagnoli_2013} and~\cite{Dodig-Crnkovic_Burgin_2011}.

In computing nature, cognition should be studied as a natural process. If we adopt the biogenetic approach to cognition, the important question is \textit{what is the minimal cognition}? Recently, a number of empirical studies have revealed an unexpected richness of cognitive behaviors (perception, information processing, memory, decision making) in organisms as simple as bacteria. Single bacteria are too small to be able to sense anything but their immediate environment, and they live too briefly to be able to memorize a significant amount of data. On the other hand bacterial colonies, swarms and films exhibit an unanticipated complexity of behaviors that can undoubtedly be characterized as biogenic cognition, \cite{Duijn_Keijzer_Franken_2006}\cite{Ben-Jacob_Shapira_Tauber_2006}\cite{Ben-Jacob_2008}\cite{Ben-Jacob_2009}\cite{Ng_Bassler_2009}\cite{Waters_Bassler_2005}.

Apart from bacteria and similar organisms without nervous system (such as e.g. slime mold, multinucleate or multicellular Amoebozoa, which recently has been used to compute shortest paths), even plants are typically thought of as living systems without cognitive capacities. However, plants too have been found to possess memory (in their bodily structures that change as a result of past events), the ability to learn (plasticity, ability to adapt through morphodynamics), and the capacity to anticipate and direct their behavior accordingly. Plants are argued to possess rudimentary forms of knowledge, according to~\cite{Pombo_Torres_Symons_2012} p. 121,~\cite{Rosen_1985} p. 7 and~\cite{Popper_1999} p. 61. 

In this article we focus on primitive cognition as the totality of processes of self-generation, self-regulation and self-maintenance that enables organisms to survive using information from the environment. The understanding of cognition as it appears in degrees of complexity can help us better understand the step between inanimate and animate matter Ð from the first autocatalytic chemical reactions to the first autopoietic proto-cells.

\section{Informational Structure of Reality for a Cognitive Agent}
\label{sec:Informational Structure of Reality for a Cognitive Agent}

When we talk about computing nature, we can ask: what is the ÒhardwareÓ for this computation? We, as cognizing agents interacting with nature through information exchange, experience nature cognitively as information. Informational structural realism ~\cite{Floridi_2003}\cite{Sayre_1976}\cite{Stonier_1997} is a framework that takes information as the fabric of the universe (for an agent). The physicists Zeilinger ~\cite{Zeilinger_2005} and Vedral ~\cite{Vedral_2010} suggest that information and reality are one. For a cognizing agent in the informational universe, the dynamical changes of its informational structures make it a huge computational network where computation is understood as information dynamics (information processing). Thus the substrate, the ``hardware'', is information that defines data-structures on which computation proceeds.

Info-computationalism is a synthesis of informational structural realism and natural computationalism (pancomputationalism) - the view that the universe computes its own next state from the previous one~\cite{Chaitin_2007}. It builds on two basic complementary concepts: information (structure) and computation (the dynamics of informational structure) as described in~\cite{Dodig-Crnkovic_2011a}~\cite{Dodig-Crnkovic_2006} and~\cite{Dodig-Crnkovic_2014}.

The world for a cognizing agent exists as potential information, corresponding to KantÕs \textit{das Ding an sich}. Through interactions, this potential information becomes actual information, ``a difference that makes a difference''~\cite{Bateson_1972}. Shannon describes the process as the conversion of latent information into manifest information~\cite{McGonigle_Mastrian_2012}. Even though BatesonÕs definition of information as a difference that makes a difference (for an agent) is a widely cited one, there is a more general definition that includes the fact that information is relational and subsumes BatesonÕs definition: 

\begin{quotation}
``Information expresses the fact that a system is in a certain configuration that is correlated to the configuration of another system. Any physical system may contain information about another physical system.'' ~\cite{Hewitt_2007} p. 293
\end{quotation}

Combining the Bateson and Hewitt insights, at the basic level, information is a difference in one physical system that makes a difference in another physical system. 

When discussing cognition as a bioinformatic process of special interest, there is the notion of agent, i.e. a system able to act on its own behalf~\cite{Dodig-Crnkovic_2014}. Agency has been explored in biological systems by~\cite{Kauffman_1995} \cite{Kauffman_1993} \cite{Deacon_2011}. The world as it appears to an agent depends on the type of interaction through which the agent acquires information,~\cite{Dodig-Crnkovic_MŸller_2011}. Agents communicate by exchanging messages (information) that help them coordinate their actions based on the (partial) information they possess (a form of social cognition).

\section{Information Self-Structuring through Morphological/Physical/Intrinsic Computation and PAC Algorithms}
\label{sec:Information Self-Structuring through Morphological/Physical/Intrinsic Computation and PAC Algorithms}

Regarding computational models of biological phenomena, we must emphasize that within the info-computational framework computation is defined as information processing. This differs from the traditional Turing machine model of computation that is an algorithm/effective procedure/recursive function/formal language. The Turing machine is a logical construct, not a physical device ~\cite{Cooper_2012}. Modeling computing nature adequately, including biological information processing with its self-generating and learning real-time properties, requires new models of computation such as interactive and networked concurrent computation models, as argued in~\cite{Dodig-Crnkovic_Giovagnoli_2013} and~\cite{Dodig-Crnkovic_2011b} with reference to~\cite{Hewitt_2012} and~\cite{Abramsky_2008}. 

Computation in nature can be described as a self-generating system consisting of networks of programs~\cite{Goertzel_1994}, a model inspired by the self-modifying systems of \cite{Kampis_1991}. In the course of the development of the general theory of networked physical information processing, the idea of computation becomes generalized. Examples of new computing paradigms include natural computing~\cite{Rozenberg_BŠck_Kok_2012} \cite{MacLennan_2004} \cite{Nunes_de_Castro_2007} \cite{Cardelli_2009}; superrecursive algorithms \cite{Burgin_2005}; interactive computing~\cite{Wegner_1998}; actor model \cite{Hewitt_2012} and similar ``second generation'' models of computing \cite{Abramsky_2008}. 

Among novel models of computation of special interest are ValiantÕs ecorythms or algorithms satisfying ``Probably Approximately Correct'' criteria (PAC) as they explicitly model natural systems ``learning and prospering in a complex world''. ~\cite{Valiant_2013} The difference between PAC learning algorithms and the Turing machine model is that the latter does not interact with the environment, and thus does not learn. It has unlimited resources, both space (memory) and time, and even though it is sequential, it does not operate in real time. In order to computationally model living nature, we need suitable resource-aware learning algorithms, such as ecorithms, described by Valiant:

\begin{quote}
``The model of learning they follow, known as the probably approximately correct model, provides a quantitative framework in which designers can evaluate the expertise achieved and the cost of achieving it. These ecorithms are not merely a feature of computers. I argue in this book that such learning mechanisms impose and determine the character of life on Earth. The course of evolution is shaped entirely by organisms interacting with and adapting to their environments.'' ~\cite{Valiant_2013} p. 8
\end{quote}

A different approach to evolution is taken by Chaitin, who argues for DarwinÕs theory from the perspective of gene-centric metabiology~\cite{Chaitin_2012}. The interesting basic idea that life is software (executable algorithms) run by physics is applied in the search for biological creativity (in the form of increased fitness). DarwinÕs idea of common descent and the evolution of organisms on earth is strongly supported by computational models of self-organization through information processing i.e. morphological computing.

The cognitive capacity of living systems depends on \textit{the specific morphology of organisms }that enables perception, memory, information processing and agency. As argued in~\cite{Dodig-Crnkovic_2012}, morphology is the central idea connecting computation and information. The process of mutual evolutionary shaping between an organism and its environment is a result of information self-organization. Here, both the physical environment and the physical body of an agent can be described by their informational structure that consists of data as atoms of information. Intrinsic computational processes, which drive changes in informational structures, result from the operation of physical laws. The environment provides an organism with a variety of inputs in the form of both information and matter-energy, where the difference between information and matter-energy is not in the kind, but in the use the organism makes of it. As \textit{there is no information without representation}~\cite{Landauer_1991}, \textit{all information is carried by some physical carrier }(light, sound, radio-waves, chemical molecules, etc.). The same physical object can be used by an organism as a source of information and as a source of nourishment/matter/energy. In general, the simpler the organism, the simpler the information structures of its body, the simpler the information carriers it relies on, and the simpler its interactions with the environment. 

\section{Cellular Computation}
\label{sec:Cellular Computation}

The environment is a \textit{resource}, but at the same time it also imposes \textit{constraints} that limit an agentÕs space of possibilities. In an agent that can be described as a complex informational structure, constraints imposed by the environment drive the time development (computation) of its structures to specific trajectories. This relationship between an agent and its environment is called \textit{structural coupling} by~\cite{Maturana_Varela_1980}. Experiments with bacteria performed by Ben-Jacob and Bassler show that bacteria interact with the environment, sense it, and extract its latent/potential information. This information triggers cognitive processes (``according to internally stored information'') that result in changes of their structure, function and behavior. Moreover, Ben-Jacob explains how information can be seen as inducing  ``an internal condensed description (model of usable information)'' of the environment, which directs its behavior and function. This is a process of intracellular computation, which proceeds via  ``gene computation circuits or gene logical elements'', that is gene circuits or regulatory pathways. As bacteria multiply by cell division, complex colony forms.

Every single bacterium is an autonomous system with internal information management capabilities: \textit{interpretation}, \textit{processing} and \textit{storage} of information. Ben-Jacob has found that complex forms emerge as a result of the communication between bacteria as interplay of the micro-level vs. macro-level (single organism vs. colony). Chemical sign-processes used by bacteria for signaling present a rudimentary form of language. Waters and Bassler~\cite{Waters_Bassler_2005} describe the process of ``quorum-sensingÓ and communication between bacteria that use two kinds of languages -- intra-species and inter-species chemical signalling. That is how they are capable of building films consisting of a variety of species.

Experiments show that the colony as a whole  ``behaves much like a multi-cellular organism'' governed by the distributed information processing with message broadcasting that stimulates changes in individual bacteria (plasticity). \textit{Communication, cooperation and self-organization} within a swarm/colony enable decision-making at the group level as a form of \textit{social cognition}.

\begin{quotation}
``The cells thus co-generate new information that is used to collectively assume newly engineered cell traits and abilities that are not explicitly stored in the genetic information of the individuals. Thus, the bacteria need only have genetically stored the guidelines for producing these capabilities.''~\cite{Ben-Jacob_2009} p. 88
\end{quotation}

A bacteria colony changes its morphology and organization through natural distributed information processing and thus learns from experience (such as encounters with antibiotics). Ben-Jacob concludes that they `` possibly alter the genome organization or even create new genes to better cope with novel challenges.'' All those processes can be modelled as distributed concurrent computation in networks of networks of programs, where individual bacteria form networks and bacteria themselves can be modelled as networks of programs (processes or executing algorithms). 

Empirical studies of the cognitive abilities of bacteria swarms, colonies and films confirm the result of~\cite{Harms_2006}, proving a theorem that natural selection will always lead a population to accumulate information, and so to 'learn' about its environment. Okasha points out that

\begin{quotation}
 ``any evolving population 'learns' about its environment, in Harms' sense, even if the population is composed of organisms that lack minds entirely, hence lack the ability to have representations of the external world at all.''~\cite{Okasha_2005}

\end{quotation}Experimental results by \cite{Ben-Jacob_Shapira_Tauber_2006}\cite{Ben-Jacob_2008}\cite{Ben-Jacob_2009}\cite{Ng_Bassler_2009}\cite{Waters_Bassler_2005} have shown that bacteria indeed learn from the environment even though the mechanisms of bacterial cognition are limited to relatively simple chemical information processes.

\section{Self-Organization, Cognitive Info-Computation and Evolution of Life}
\label{sec:Self-Organization, Cognitive Info-Computation and Evolution of Life}

In computational (information processing) models of bacterial cognition, the biological structure (hardware) is at the same time a program (software) that controls the behavior of that hardware both internally and in the interactions with the environment. Already in 1991 Kampis proposed a unified model of computation as the mechanism underlying biological processes through Òself-generation of information by non-trivial change (self-modification) of systemsÓ ~\cite{Kampis_1991}. This process of self-organization and self-generation of information is what is elsewhere described as morphological computation on different levels of organization of natural systems. Current research in adaptive networks goes in the same direction,\cite{Dodig-Crnkovic_Giovagnoli_2013}.

However, understanding of the basic evolutionary mechanisms of information accumulation, with resulting increase in information-processing capacities of organisms (memory, anticipation, computational efficiency), is only the first step towards a fully-fledged evolutionary understanding of cognition, though it is probably the most difficult one, as it requires a radical redefinition of fundamental concepts of information, computation and cognition in naturalist terms. According to Maturana:

\begin{quotation}
``A cognitive system is a system whose organization defines a domain of interactions in which it can act with relevance to the maintenance of itself, and the process of cognition is the actual (inductive) acting or behaving in this domain. Living systems are cognitive systems, and living as a process is a process of cognition. This statement is valid for all organisms, with and without a nervous system.'' ~\cite{Maturana_1970} p. 13 
\end{quotation}

The role of cognition for a living agent, from bacteria to humans is \textit{to efficiently deal with the complexity of the world}, helping an agent to survive and thrive. The world is inexhaustible and largely complex and exceeds by all accounts what a cognizing agent can take in. Cognition is then the mechanism that enables cognizing agents to control their own behavior in order to deal with the complexity of the environment, make sense of the world and use it as a resource for survival, \cite{Godfrey-Smith_2001} p. 234. In this view, `` cognition ` shades off' into basic biological processes such as metabolism.''

Through autopoietic processes with structural coupling (interactions with the environment) a biological system changes its structures and thereby the information processing patterns in a self-reflective, recursive manner~\cite{Maturana_Varela_1980}. But self-organisation with natural selection of organisms, responsible for nearly all information that living systems have built up in their genotypes and phenotypes, is a simple but costly method to develop. Higher organisms (which are Òmore expensiveÓ to evolve in terms of resources) have developed language and reasoning as a more efficient way of learning. The step from Ògenetic learningÓ (typical of more primitive forms of life) to the acquisition of cognitive skills on higher levels of organisation of the nervous system (such as found in vertebrata) will be the next step to explore in the project of cognitive info-computation, following~\cite{Jablonka_Lamb_2005} who distinguish \textit{genetic, epigenetic, behavioral, and symbolic evolution}. The studies of bacterial cognition suggest that there are some important processes that operate during evolution such as self-organization and auto-poiesis, which guarantee growth of order, and the propagation of structures in spite of the randomness of environmental influences. Also, colonies, swarms and films seem to play a prominent role in bacterial evolution (as Òswarm intelligenceÓ).

Interesting question arises in connection to AI and AL which are not based on chemical processes: \textit{is molecular computation necessary for cognition?} For example \cite{Duijn_Keijzer_Franken_2006} proposed that minimal cognition can be identified with sensorimotor coordination. However, even though fundamental, sensorimotor coordination is not enough to explain cognition in biological systems. Chemical processes of autopoiesis based on molecular computation (information processing) are essential, not only for simple organisms like bacteria, but also for the functioning of the human nervous system. In the words of Shapiro:

\begin{quotation}
``molecular biology has identified specific components of cell sensing, information transfer, and decision-making processes. We have numerous precise molecular descriptions of cell cognition, which range all the way from bacterial nutrition to mammalian cell biology and development.'' \cite{Shapiro_2011} p. 24
\end{quotation}

Info-computational approach provides an appropriate framework for studying the above question of minimal cognition. The advantages of info-computational approaches to the modeling of cognition are that they bridge the Cartesian gap between matter and mind, providing a unified naturalist framework for a vast range of phenomena, and they are testable. Dennett declared in a talk at the International Computers and Philosophy Conference, Laval, France in 2006: ``AI makes philosophy honest.'' Paraphrasing Dennett we can say that info-computational models make cognition honest - - transparent and open for critical investigation and experimentation. In that sense parallel research in biology and cognitive robotics present a ``reality check'' where our understanding of cognition, information processing and morphological computation can be tested in a rigorous manner. 

\section{Conclusions}
\label{sec:Conclusions}

Studied as a natural phenomenon, cognition can be seen as info-computational processes in living systems. The aim of this article is to present methodological and practical grounds for a naturalist computational approach to cognition supported by new experimental results on cognition of simplest living organisms such as bacteria. The hope is to contribute to the elucidation of the following fundamental questions according to~\cite{Bechtel_1998}~\cite{Lyon_2005} and~\cite{Duijn_Keijzer_Franken_2006}:

\textit{What cognition is.} The nature of cognition, the question about how the concept of cognition should be defined. In the info-computational framework it becomes transformed into the question: \textit{what in the computing nature is cognition?} Cognition for an adaptive, developing and evolving living agent is the process of learning that operates according to the PAC (Probably Approximately Correct) strategy ~\cite{Valiant_2013}. Results from the studies of natural cognitive systems will help resolve the question concerning artifactual computational cognition.

\textit{How cognition works.} Cognition as information processing happening in an informational network of cognizing agents with distributed computational dynamics connects the agentÕs inside structures with the outside world understood as potential information, through interactions. Those interactions include all four levels on which evolution operates: \textit{genetic, epigenetic, behavioral, and symbolic} ~\cite{Jablonka_Lamb_2005}. We have shown in the example of bacterial cognition how all four levels contribute.

\textit{What cognition does.} By elucidating the info-computational and evolutionary character of cognition we can understand its agent-dependency, its generative mechanisms and its life-sustaining and life-propagating role. Cognition is the mechanism that enables cognizing agents to deal with the complexity of the environment, through control of their own behavior,~\cite{Godfrey-Smith_2001} p. 234. 

The info-computational approach can contribute to rethinking cognition as information self-organising processes of morphological/chemical/molecular/natural computation \textit{in all living beings}. Thus, we can start to learn how to adequately computationally model living systems, which has up to now been impossible,~\cite{Dodig-Crnkovic_MŸller_2011}. ``Second generation computational models''~\cite{Abramsky_2008} under current development promise to enble us to frame theoretically, simulate and study living organisms in their full complexity. Based on current work in the related fields such as information science, computability and theory of computing, logic, molecular biology, and evolution, a new more coherent picture of cognition can be expected to emerge. As a complement to WoframÕs idea of \textit{Òmapping and mining the computational universeÓ}~\cite{Wolfram_2002} this article suggests \textit{mapping and mining the biological universe} with the help of computational tools with the goal to reverse engineer cognition and find smart cognitive computational strategies.

\bibliographystyle{splncs}
\bibliography{Bibliography-CiE2014}
\end{document}